\documentstyle[12pt]{article}

\large
\newcommand{\be}{\begin{eqnarray}}
\newcommand{\ee}{\end{eqnarray}}
\newcommand\del{\partial}

\newcommand\dg{\dagger}

\begin{document}
\setlength{\baselineskip}{21pt}
\pagestyle{empty}
\vfill
\eject
\begin{flushright}
SUNY-NTG-96/29
\end{flushright}

\vskip 2.0cm
\centerline{\large \bf Finite volume partition functions }
\centerline{\large\bf and Itzykson-Zuber integrals}
\vskip 2.0 cm
\centerline{A.\,D. Jackson, M.\,K. \c Sener, J.\,J.\,M. Verbaarschot}
\vskip 0.2cm
\centerline{Department of Physics}
\centerline{SUNY, Stony Brook, New York 11794}
\vskip 2cm

\centerline{\bf Abstract}
We find the finite volume QCD partition function for different quark
masses. This is a generalization of a result obtained by Leutwyler and
Smilga for equal quark masses.  Our result is derived in the sector of
zero topological charge using a generalization of the Itzykson-Zuber
integral appropriate for arbitrary complex matrices.  We present a
conjecture regarding the result for arbitrary topological charge which
reproduces the Leutwyler-Smilga result in the limit of equal quark
masses.  We derive a formula of the Itzykson-Zuber type for arbitrary
{\em rectangular\/} complex matrices, extending the result of Guhr and
Wettig obtained for {\em square\/} matrices.

\vfill
\noindent
\begin{flushleft}
May 1996
\end{flushleft}
\eject
\pagestyle{plain}

\noindent
\section{Introduction}
\vskip 0.5cm

In general, numerical simulations offer our only access to the QCD partition
function.  For sufficiently small volumes, however, the partition function
is dominated by the constant modes which makes analytic treatment possible.
(See, {\em e.g.}, ref.\,\cite{vanbaal}.)  In a similar spirit, Leutwyler
and Smilga \cite{LS} identified a parameter range within which the mass
dependence of the partition function is completely determined by the
underlying structure of broken chiral symmetry.  This range is given as
\be
\frac 1{\Lambda} \ll L \ll \frac 1{m_{\pi}} \ ,
\label{range}
\ee
where $L$ is the linear size of the 4-dimensional Euclidean box, $\Lambda$
is a typical hadronic mass scale and $m_\pi$ is the mass of the Goldstone
modes ({\em i.e.}, $m_\pi \sim \sqrt {m \Lambda}$ for quark mass $m$).
The lower limit of this range ensures that the partition function is
dominated by the Goldstone modes.  The upper limit ensures that these modes
can be treated as constant modes.  If chiral symmetry is broken according
to $SU(N_f)\times SU(N_f) \rightarrow SU(N_f)$, the QCD partition function
in the range (\ref{range}) and for vacuum angle $\theta$ is given by \cite{LS}
\be
Z(M,\theta) = \int_{U\in SU(N_f)} dU \exp\left(V\Sigma\,
{\rm Re}(e^{i\theta/N_f}\,{\rm tr} \; U M)\right) \ .
\label{ztheta}
\ee
Here, $\Sigma$ is the chiral condensate, and $M$ is the mass matrix which
can be taken as diagonal without loss of generality.  In the sector with
topological charge $\nu$, the partition function is given as
\be
Z_{\nu}(M) = \frac 1{2\pi} \int_0^{2\pi}d\theta  e^{-i\nu \theta} Z(\theta) \ .
\label{znu}
\ee

For equal quark masses, Leutwyler and Smilga obtained an exact analytic
expression for this partition function
\be
Z_\nu(m) = \det_{ij} I_{\nu +j-i}(m\,V\,\Sigma) \ ,
\label{ZLS}
\ee
where $i$ and $j$ run from $1, \cdots, N_f$.  The importance of this
partition function lies in the fact that it enables us to determine
the volume dependence of the chiral condensate.  This is of interest
in lattice QCD simulations, where the volume is necessarily finite.

Recently, lattice QCD calculations have been performed to determine
the connected and disconnected contributions to the scalar susceptibility
\cite{Karsch}.  The determination of these susceptibilities requires
differentiation of the partition function with respect to two
{\em different\/} masses. In order to determine volume dependence of
the scalar susceptibility in the range (\ref{range}), we require a
generalization of (\ref{ZLS}) to different quark masses.  This is the
primary objective of the present paper.

In section 2, we analyze the two flavor case and generalize the result
to an arbitrary number of flavors. In section 3, we present a generalization
of the Itzykson-Zuber formula valid for arbitrary {\em rectangular\/}
complex matrices.  For {\em square\/} complex matrices, our result reduces
to that obtained by Wettig and Guhr \cite{Wettig-Guhr}.  In section 4, we
derive the finite volume partition function for different masses in the
sector of zero topological charge.  We also make a conjecture of the result
for arbitrary topological charge and show that it reduces to (\ref{ZLS})
for the special case of equal quark masses.

\vskip 1.5cm
\noindent
\section{The finite volume partition function for $N_f =2$ and its
generalization to arbitrary $N_f$.}
\vskip 0.5cm

For two flavors and in the range (\ref{range}), the QCD-partition function
is known for different quark masses.  For vacuum angle $\theta$, it is given
as
\be
Z(\theta) = \frac 2{V \, \Sigma \, \mu} I_1(V\,\Sigma\,\mu) \ ,
\ee
where $I_1$ is a modified Bessel function and where
\be
\mu^2 = m_1^2 +m_2^2 + 2m_1m_2 \cos(\theta) \ .
\ee
The partition function in the sector with topological charge $\nu$
can be obtained by integrating over $\theta$ according to (\ref{znu}).
Remarkably, this integral can be expressed analytically.  After some
manipulations the result can be written as
\be
Z_\nu(N_f=2) =  \frac 2{x_2^2-x_1^2} \det \left|
\begin{array}{cc}I_\nu(x_1) & x_1 I_\nu'(x_1)\\ I_\nu(x_2) & x_2 I_\nu'(x_2)
\end{array} \right|
\ee
where
\be
x_k = m_k V \, \Sigma.
\ee
This suggests the generalization to three flavors
\be
Z_\nu(N_f = 3) =\frac {16}{(x_2^2-x_1^2)(x_2^2-x_3^2)(x_3^2-x_1^2)} \det \left|
\begin{array}{ccc}I_\nu(x_1) & x_1 I_\nu'(x_1)& x_1^2 I_\nu''(x_1)
\\ I_\nu(x_2) & x_2 I_\nu'(x_2) &x_2^2 I_\nu''(x_2)
\\ I_\nu(x_3) & x_3 I_\nu'(x_3) &x_3^2 I_\nu''(x_3)
\end{array} \right| \ \ ,
\ee
where the numerical prefactor is chosen such that for equal masses the
result of Leutwyler is reproduced.
The generalization to an arbitrary number of flavors is now obvious.  We
define a Vandermonde determinant as
\be
\Delta(x^2) = \prod_{k<l} \, (x_k^2 - x_l^2) \ ,
\label{vandermonde}
\ee
and an $N_f \times N_f$ matrix as
\be
A_{kl} = x_k^{l-1} I_\nu^{(l-1)}(x_k ) \qquad k,l = 1, \cdots, N_f.
\label{Anukl}
\ee
The partition function is then given as
\be
Z_\nu(m_1, \cdots, m_{N_f}) = C_{N_f}\frac{\det A}{\Delta(x^2)} \ ,
\label{Znu1}
\ee
where the normalization constant
\be
C_{N_f} = 2^{N_f(N_f-1)/2} \prod_{k=1}^{N_f} (k-1)!
\ee
is determined by the limit of equal quark masses.

In section 4, we shall prove this formula for the special case $\nu = 0$.
It will be shown that this formula reduces to the Leutwyler-Smilga finite
volume partition function for arbitrary $\nu$.

\section{The Itzykson-Zuber integral for complex rectangular matrices}

In this section we offer a derivation of the extension of the Itzykson-Zuber
integral to the case of arbitrary complex {\em rectangular\/} matrices using
the diffusion equation method.  The result for square matrices has also been
obtained in \cite{Wettig-Guhr}.  Our derivation is patterned on the argument
for Hermitean matrices, which has been discussed widely in the literature.
(See, {\em e.g.}, \cite{Mehta81,Guhr91,Ginsparg}.)  The expression which will
be required in section 4 for the calculation of the finite volume partition
function for different masses is given in (\ref{IZ2}).

\medskip

Let $\sigma$ and $\rho$ be arbitrary complex ({\em i.e.}, non-Hermitian) $N_1
\times N_2$ matrices.  Without loss of generality, we assume that
$\nu \equiv N_1 - N_2 \ge 0$.  Any such matrix can be diagonalized
in the form \cite{Hua}
\be
\sigma =  U^\dg S V
\label{diagonal}
\ee
where
\be
S =  \left(
            \begin{array}{c}
                \hat{S}\\
                {\bf 0}
            \end{array}
     \right) \ ,
\ee
and where the square diagonal matrix
$\hat{S} = {\rm diag}(s_1,\cdots, s_{N_2})$
has nonnegative real entries. The matrices $(U,V)$ parameterize the coset
space $U(N_1) \times U(N_2)/[U(1)]^{N_2}$.  Similarly, $\rho$ can be written
as $\rho = {U'}^\dg R V'$.

In the following, we evaluate the integral
\be
 \frac 1{ (\pi t)^{ (N_1 N_2)}}
 \int d\mu(U,V) \ {\rm exp}\left( -\frac{1}{t}
 {\rm tr} [ (\sigma - \rho)^\dg (\sigma - \rho) ] \right)
\ee
using the diffusion equation method and exploiting the invariance of the
measure, $d\mu(U,V)$, which is taken to be the Haar measure of
$U(N_1) \times U(N_2)/[U(1)]^{N_2}$.
The function
\be
F(\rho,t) =  \frac 1{ (\pi t)^{ (N_1 N_2)}}
\int d[\sigma] \, {\rm exp}\left( -\frac{1}{t}
 {\rm tr} [ (\sigma - \rho)^\dg (\sigma - \rho) ] \right) \varphi(\sigma)
\label{Frho}
\ee
with integration measure $d[\sigma] =
\prod_{m=1}^{N_1} \prod_{n=1}^{N_2}
d {\rm Re}\sigma_{mn}  d {\rm Im}\sigma_{mn}$
satisfies the diffusion equation
\be
\sum_{m=1}^{N_1} \sum_{n=1}^{N_2}
\frac {\del^2} {\del \rho_{mn} \del \rho_{mn}^* } F(\rho,t)
=  \frac{\del }{\del t} F(\rho,t) \ .
\label{F_de}
\ee
As initial condition, we choose an invariant function, {\em i.e.},
$\varphi(\sigma)$ is a function of the eigenvalues of $\sigma$ only.
Then, using the invariance of the measure, we find immediately that
$F(\rho,t)$ is a symmetric function of the eigenvalues of $\rho$.

In order to proceed, we express both the Laplacian and the integration
measure in the `polar coordinates' introduced by the diagonalization
(\ref{diagonal}),
\be
d[\sigma] = \Omega d[S] \, j^2(S) \, d\mu(U,V)
\ee
where $d[S] = \prod_{n=1}^{N_2} ds_n$.  Here, the constant $\Omega$, which
depends on the convention adopted for the measure of the group, will be
fixed later.  The Jacobian is given by $J(S) \equiv j^2(S)$ with
\be
j(S) = \prod_{n=1}^{N_2} s_n^{(N_1 - N_2) + 1/2} \Delta(\hat{S}^2) \ ,
\label{jj}
\ee
and $\Delta(\hat S^2)$ is the Vandermonde determinant defined in
(\ref{vandermonde}).  Because $F$ is an invariant function, it satisfies
a diffusion equation which involves only the radial part of the Laplacian:
\be
\sum_{n=1}^{N_2} \frac{1}{J(R)} \frac{\del}{\del r_n} J(R)
\frac{\del}{\del r_n}F = 4 \frac {\del F}{\del t} \ \ .
\label{Fde_r}
\ee
This equation can be simplified materially with the introduction of
a reduced wave function,
\be
f(R,t) =  j(R) F(R,t) \ \ .
\label{trick}
\ee
Now, (\ref{Fde_r}) reduces to
\be
\sum_{n=1}^{N_2} \left(  \frac{\del^2}{\del r_n^2} -
\frac{1}{j(R)}  \left[ \frac{\del^2}{\del r_n^2}
j(R) \right]  \right) f(R,t) = 4 \frac{\del f}{\del t} \ \ .
\label{inbetween}
\ee
Remarkably, this differential equation is separable.\footnote{In
\cite{Wettig-Guhr}, a separable equation was obtained for $N_1 = N_2$
by the substitution of $\Delta(R^2) F(R,t)$ instead of $j(R) F(R,t)$.}
Performing the differentiations of $j(R)$, it can be rewritten as
\be
\sum_{n=1}^{N_2} \left( \frac{\del^2}{\del r_n^2} - \frac{4\nu^2 -1}{4}
 \frac{1}{r_n^2}  \right) f(R,t) = 4 \frac{\del f}{\del t} \ .
\label{bess}
\ee
Because of the presence of the factor $j(R)$ in (\ref{trick}), $f(R,t)$
is an antisymmetric function of the eigenvalues.  Therefore, the solution
of (\ref{bess}) is given by an integral over a Slater determinant
\be
f(R,t) = \int d[S] \, \frac{1}{N_2!}
\det_{k,l}\left| g(r_k,s_l;t)\right| \, j(S)
\, \varphi(S) \ ,
\label{fdet}
\ee
where $g(r,s;t)$ is the kernel of
\be
\frac{\del^2}{\del r^2} g - \frac{4\nu^2 - 1 }{4 r^2} g =
4 \frac{\del g}{\del t} \ \ .
\label{gde}
\ee
The kernel can be expressed in terms of the regular eigenfunctions of
the Bessel equation of order $\nu$:
\be
 u'' + [k^2 - (4\nu^2-1)/4r^2] u = 0 \ \ .
\label{bessnu}
\ee
This is obtained following a separation of variables which leads to a time
dependence of the form $\exp(-k^2t/4)$.  The regular eigenfunctions are
given by
\be
u_k(r) \sim \sqrt{k r} \, J_{\nu}(k r) \ ,
\ee
where $J_{\nu}$ is a Bessel function.  Recalling the orthogonality relation
for Bessel functions, the kernel of (\ref{gde}) can be written as
\be
g(r,s;t) = \theta(t) \int_0^\infty dk \, k \, \sqrt{rs}
{\rm e}^{-k^2 t/4} \, J_\nu(k r) J_\nu(k s) \ \ .
\ee
This can be evaluated to give
\be
g(r,s;t) = \theta(t) \, \frac{2}{t} \sqrt{rs}
\exp{\left( -\frac{r^2+s^2}{t} \right)} I_\nu \left( \frac{2rs}{t} \right)
\label{grst}
\ee
where $I_\nu$ is a modified modified Bessel function.

Finally, we equate the definition of $F(\rho,t)$ in (\ref{Frho}) with
its expression in terms of the kernel (\ref{grst}) as given by
(\ref{trick}) and (\ref{fdet}).  Since this equality is valid for
an arbitrary choice of the initial condition $\varphi(\sigma)$,
the integrands of $d[S]$ must be the same.

Hence, we arrive at
\be
\int d\mu(U,V) \, \exp{\left( -\frac{1}{t} {\rm tr} [ (\sigma - \rho)^\dg
(\sigma - \rho) ] \right)} \nonumber
\ee
\be
= \frac{t^{N_1 N_2 - N_2}}{\prod_{k=1}^{N_2} (r_k s_k )^{\nu}}
\frac{2^{N_2} \pi^{N_1 N_2}}{\Omega} \frac{1}{N_2!}
\det_{k,l} \left| \exp{\left( - \frac{r_k^2+s_l^2}{t} \right)
I_\nu \left( \frac{2 r_k s_l}{t} \right)} \right|/\Delta(S^2) \Delta(R^2) \ .
\label{IZ1}
\ee
Here, the value of $\Omega$ follows from the normalization integral calculated
in \cite{NW}:
\be
\Omega = \frac{\pi^{N_1 N_2} 2^{N_2}}{\prod_{j=1}^{N_2} j! (j + \nu- 1)!} \ ,
\ee
where we have used the convention that $\int d\mu(U,V) = 1$.

This result enables us to calculate the Itzykson-Zuber integral for
arbitrary complex matrices
\be
\int d\mu(U,V) \, \exp{\left( {\rm Re} \, {\rm tr} \, U^\dg S V R \right)}
= C_{N_1} C_{N_2} \frac{2^{-\nu(\nu+1)}}{\prod_{k=0}^{\nu-1} k!
\prod_{k=1}^{N_2} (r_k s_k )^{\nu}}
\det_{k,l} \left| I_\nu(r_k s_l)\right| /\Delta(S^2) \Delta(R^2) .\nonumber\\
\label{IZ2}
\ee
When $\nu=0$, the product of factorials in this expression is understood
to be 1.  The constants $C_{N_1}$ and $C_{N_2}$
can be evaluated to be
\be
C_n = 2^{n(n-1)/2} \prod_{k=1}^{n} (k-1)! \ \ .
\label{normalization}
\ee
In the special case $N_1=N_2$ for which $\nu=0$, our expression reduces to
the result of Guhr and  Wettig \cite{Wettig-Guhr} apart from a
normalization constant.

\section{The finite volume partition function for different masses}

The finite volume effective partition function of QCD in the sector
with topological charge $\nu$, defined in (\ref{znu}) and (\ref{ztheta}),
can be written as an integral over $U(N_f)$ instead of $SU(N_f)$ \cite{LS}
\be
Z_\nu = \int_{U(N_f)}
d\mu(U) (\det U)^{\nu} \exp \left(
V \Sigma \; {\rm Re} \; {\rm tr} M U^\dg \right) \ \ .
\label{ZnuLS}
\ee
For $\nu=0$, this result can be obtained from (\ref{IZ2}) by taking
$R = V \Sigma M$ and $S = {\bf 1}_{N_f}$. Because of the singularity as
$S \rightarrow {\bf 1}$, the final result requires a careful analysis of
this limit.

\medskip

In (\ref{IZ2}), we take $S={\bf 1} + \delta S$ and expand the
modified Bessel functions to order $(\delta S)^{N_f}$.
The result is
\be
I_{\nu}(r_k s_l) = \sum_{j=1}^{N_f} \, \frac {r_k^{j-1}}{(j-1)!} \,
I_{\nu}^{(j-1)}(r_k) \, \delta s_l^{j-1} \ + \ {\cal O}(\delta S^{N_f})\ ,
\ee
which can be written as the product of the matrix $A$ defined in (\ref{Anukl})
and the matrix
\be
B_{jl} = \frac{1}{(j-1)!} \, \delta s_l ^{j-1} \ .
\label{Bjl}
\ee
The determinant of $B$ can be written as
\be
\det (B) = \frac 1{\prod_{k=1}^{N_f}(k-1)!}
\Delta(\delta S) = \frac 1{C_{N_f}}
 \Delta \left( (1 + \delta S)^2 \right) \ ,
\label{detb}
\ee
with the normalization constant defined in (\ref{normalization}).
Hence
\be
Z_{\nu}(M) = C_{N_f} \frac{\det A}{\Delta(R^2)} \ ,
\label{Z0M}
\ee
which is simply the result conjectured in (\ref{Znu1}) above.  This result is
now proved for $\nu = 0$.

We have not proved the result (\ref{Znu1}) for arbitrary $\nu$.  However,
we offer one non-trivial check of this conjecture by demonstrating that
(\ref{Znu1}) reduces to the finite volume partition function of Leutwyler
and Smilga in the limit of equal masses.  We start with the expression
\be
A_{kj} = C_{N_f}^{1/N_f} r^{j-1}_k I_\nu^{(j-1)}(r_k)
\label{Akl2}.
\ee
Using the recursion relation
\be
\frac{dI_{\nu}}{dr} = I_{\nu+1} + \frac{\nu}{r} I_{\nu}
\label{rr1}
\ee
and adding a suitable multiple of the column to the left of the column
in question, we arrive at the matrix
\be
A_{kj} \rightarrow C_{N_f}^{1/N_f} r^{j-1}_k I_{\nu+j-1}(r_k) \ \ .
\ee
This is evidently correct in going from the first to the second column and,
hence, true in general.  The fact that the coefficient of $I_{\nu+1}$ in
(\ref{rr1}) is 1 guarantees that the determinant will not be affected by
this rearrangement.

To realize the limit $r_k \rightarrow r\equiv mV\Sigma$,
we write $r_k = r +\delta r_k$ and
expand each element in $A$ in a Taylor series through order $\delta
r_k^{N_f-1}$.  The result is that
\be
A_{kj} = C_{N_f}^{1/N_f}
\sum_{p=1}^{N_f}\frac 1{(p-1)!} \frac {d^{p-1}}{dr^{p-1}}
\left[ r^{j-1} I_{\nu+j-1}(r)\right] \, \delta r_k^{p-1} \ .
\ee
This can be written as the product of the matrix with elements
\be
M_{jp}= C_{N_f}^{1/N_f}\frac {d^{p-1}}{dr^{p-1}} \, \left[
r^{j-1} \, I_{\nu+j-1}(r) \right]
\ee
and the matrix $B_{jp}\equiv (\delta r_k)^{p-1}/(p-1)!$.  As in (\ref{detb}),
we have
\be
\det(B) = \prod_{p=1}^{N_f} \frac 1{(p-1)!}
\Delta(\delta R) = \frac 1{C_{N_f}}\left( \frac{1}{r} \right )^{N_f(N_f-1)/2}
\ \ \Delta((R+\delta R)^2) \ .
\ee
The determinant of the matrix $M$ can be simplified by using the
recursion relation
\be
\frac{dI_{\nu}}{dr} = I_{\nu-1} - \frac{\nu}{r} I_{\nu}
\label{rr2}
\ee
and adding a suitable multiple of the row immediately above the row in
question.  As before, this rearrangement does not alter the determinant.
This leaves us with
\be
M_{kj} \rightarrow  C_{N_f}^{1/N_f}r^{j-1} I_{\nu+j-k} \ .
\label{Afinal}
\ee
Now, the factor $r^j$ can be extracted from each column and used to eliminate
the $r$ dependence in the prefactor. Thus, we arrive at the final result
\be
\det (A) = \det(M) \det(B) = \det_{k,j}I_{\nu+j-k} \ ,
\ee
which is precisely the result in \cite{LS} as given
in (\ref{ZLS}).

\noindent\section{Conclusions}
\vskip 0.5cm
We have obtained the finite volume QCD partition function for different
quark masses in the range $1/\Lambda \ll L \ll 1/m_{\pi}$.  This result
generalizes the finite volume partition function obtained previously
by Leutwyler and Smilga for the case of equal quark masses. In order
to derive this result, we were led to generalize the Itzykson-Zuber
integral to arbitrary rectangular complex matrices. The integral for
square matrices, first obtained by Guhr and Wettig, leads immediately
to the proof in the sector of zero topological charge. Based on the
result for two flavors and the general result for $\nu = 0$, we have
conjectured the result for arbitrary $\nu$ and $N_f$.  As a decidedly
nontrivial check of this conjecture, we have shown that the result of
Leutwyler and Smilga emerges in the limit of equal quark masses.

We wish to note a remarkable coincidence.  Consider the Itzykson-Zuber
formula for complex rectangular matrices (with $\nu$ equal to the
difference between the number of rows and columns) in the same limit
considered for $\nu=0$.  Up to a factor,
this leads to the finite-volume partition
function in the sector of topological charge $\nu$.  While we can offer
no explanation of this coincidence, it may be useful to note that,
in the chiral limit, the joint eigenvalue density of the random matrix
model associated with the finite volume partition function depends
only on the combination $\nu + N_f$ \cite{V}.  Clearly, more work lies
ahead in this area.

\vglue 0.6cm
{\bf \noindent  Acknowledgements \hfil}
\vglue 0.4cm
 The reported work was partially supported by the US DOE grant
DE-FG-88ER40388.

\end{document}